\documentclass[prb,preprint,aps]{revtex4-1}
\usepackage{graphicx}
\usepackage{epstopdf}
\newcommand{\dt}{{\Delta t}}

\newcommand{\p}{{\bf p}}
\newcommand{\bv}{{\bf v}}
\newcommand{\ac}{{\bf a}}

\newcommand{\br}{{\bf r}}

\newcommand{\bea}{\begin{eqnarray}}
\newcommand{\eea}{\end{eqnarray}}
\newcommand{\be}{\begin{equation}}
\newcommand{\ee}{\end{equation}}
\newcommand{\ba}{\begin{eqnarray}}
\newcommand{\ea}{\end{eqnarray}}

\newcommand{\nn}{\nonumber}
\newcommand{\la}{\label} 
\newcommand{\w}{\omega} 
\newcommand{\W}{\Omega}

\def\t1{e_{_T}}
\def\v1{e_{_V}}

\begin{document}
\title{Modern light on ancient feud: Robert Hooke and Newton's graphical method}

\author{Siu A. Chin}

\affiliation{Department of Physics and Astronomy, 
Texas A\&M University, College Station, TX 77843, USA}

\begin{abstract}
	
	The feud between Robert Hooke and Isaac Newton has
remained ongoing even after 300 years, over whether Newton should have 
acknowledged Hooke's influence on his graphical method of
constructing planet orbits, the celebrated Proposition 1, 
Theorem 1 of the {\it Principia}. The drama has escalated in
recent decades, with a claim that Hooke may have used the same 
method and obtained an elliptical orbit for a linear force,
a feat that some considered Newton never did for the inverse-square force.
Modern understanding of Newton's graphical method as a 
symplectic integrator can now shed light on whether this claim is creditable. 
This work, based on knowing the Hamiltonian of the symplectic integrator, 
deduced the analytical orbit corresponding to Newton's graphical 
construction. A detailed comparison between
this analytical orbit and Hooke's drawing shows that it is unlikely
that Hooke had used Newton's graphical method and obtained the 
correct orbit.

\end{abstract}
\maketitle

\section {Introduction}
\la{intro}
    
    Historians of science have continued to debate Robert Hooke's
influence on Newton's formulation of orbital 
dynamics\cite{loh60,wes67,cen70, hun89,pug89,gal02,coo03, hun06,pur09}. 
According to a series of well-studied correspondences\cite{tur60}, Hooke wrote a letter
to Newton on November 24, 1679, asking for his thought on Hooke's hypothesis
of ``compounding" the ``direct motion by the tangent" of the planet and 
the ``attractive motion" toward the central body. Newton replied that he did not
hear about Hooke's hypothesis prior to this letter, a claim that some have viewed 
as ``evasive"\cite{bar01}, ``disingenuous"\cite{coo03a}, or even possibly a ``bare-face lie"\cite{gal02a}. 
In 1684 Newton deposited the {\it De Motu}, an initial draft of the {\it Principia}, 
with the Royal Society, which Hooke has access. 
In the {\it De Motu}, Newton showed a graphical construction 
of a central force orbit incorporating Kepler's equal area law, which 
later became Proposition 1, Theorem 1 of 
the {\it Principia}. (See Fig.\ref{prop1}.) To Hooke, and to many of his later supporters\cite{cen70,hun89,pug89,gal02,coo03,hun06,pur09,nau94}, 
this graphical construction seemed the exact embodiment of Hooke's idea of 
compounding tangential inertial motion with a central 
attractive force, yet Newton gave Hooke no credit whatsoever. 

\begin{figure}
	\includegraphics[width=0.70\linewidth]{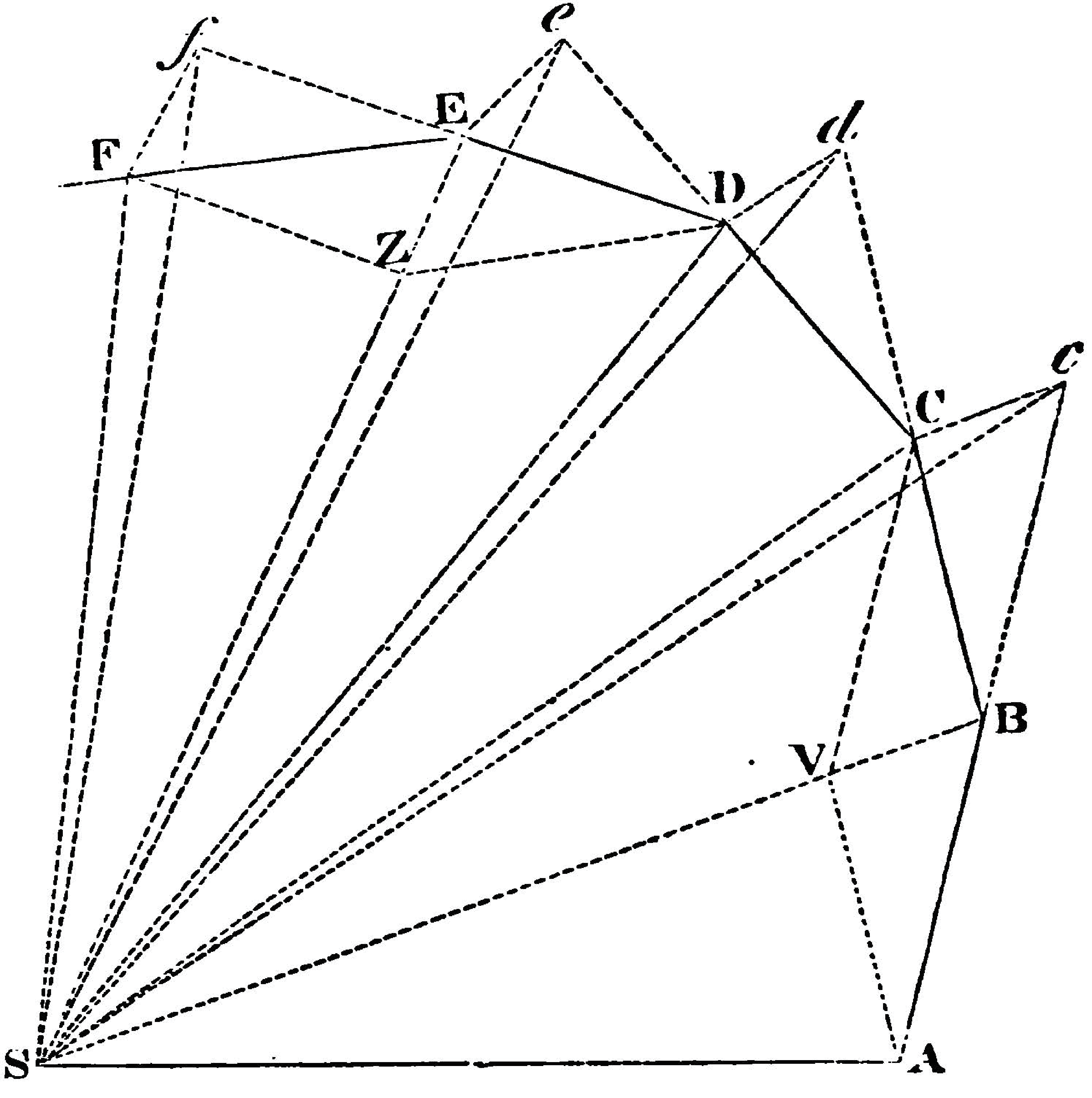}
	\caption{Newton's iconic graphical construction; the {\it Principia}'s Proposition 1,
		Theorem 1, taken from Ref.\onlinecite{prop1} and further explained in the text.	
	} 
	\la{prop1}
\end{figure} 

Newton's action was probably due to his strong and negative reaction to Hooke's interim 
accusation of him plagiarizing the inverse-square law. 
A year before the publication of 
the {\it Principia}, Edmond Halley wrote to Newton on May 22, 1686, relating
that ``... Mr Hooke has some pretension upon the invention of the rule
of decrease of gravity being reciprocally as the squares of the distances 
from the center. He says you had the notion from him, ... Mr Hook seems 
to expect you should make some mention of him, in the preface 
[of the {\it Principia}]..."\cite{tur60} Newton was furious and wrote back on 
June 20, 1686 that ``he [Hooke] has done nothing and yet
written in such a way as if he knew and had sufficiently hinted all but what 
remained to be determined by the drudgery of calculations and observations, 
excusing himself from that labour by reason of his other business: whereas 
he should be rather have excused himself by reason of his inability."\cite{tur60}

\begin{figure}
	\centering
	\includegraphics[width=0.80\linewidth]{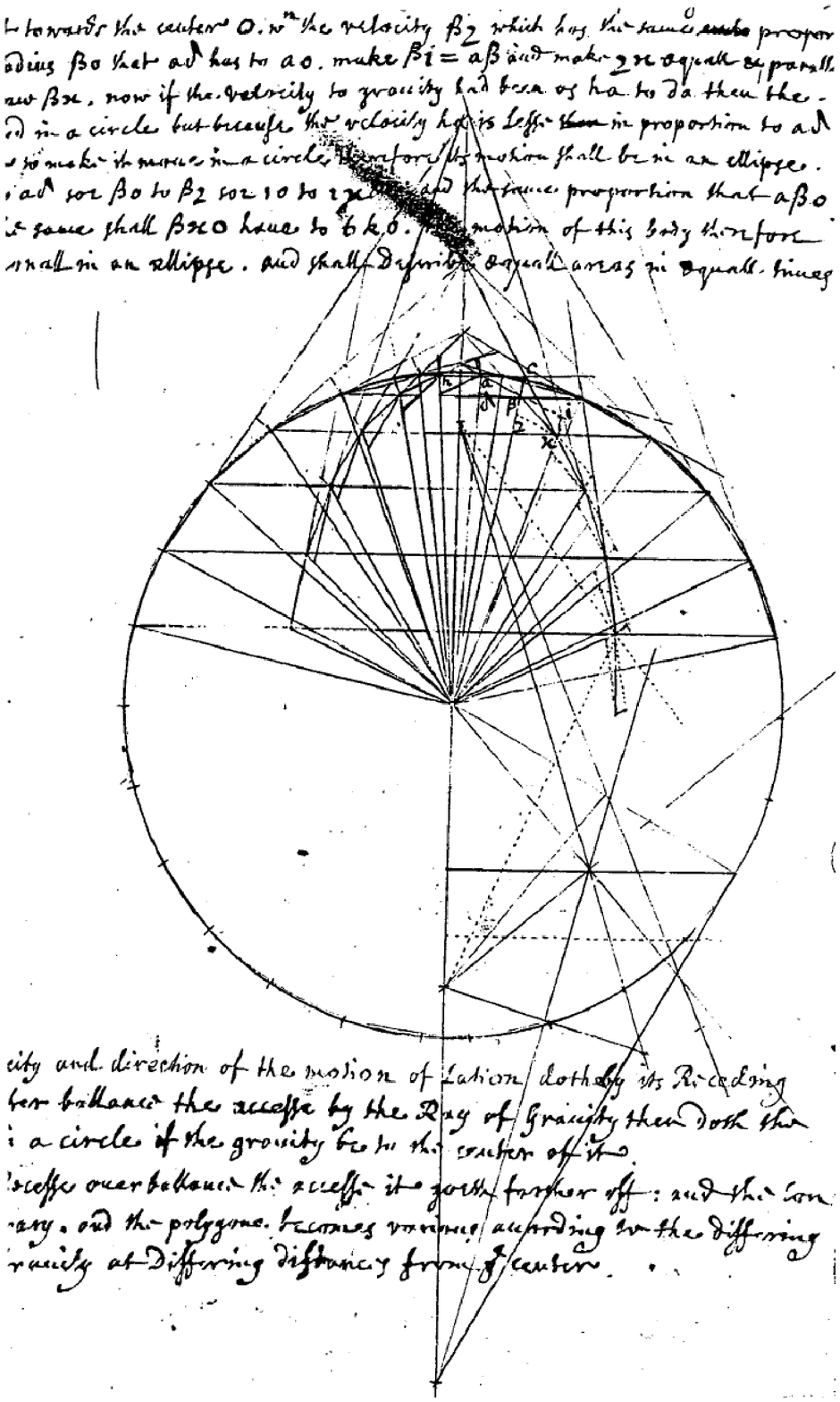}
	\caption[]{Hooke's unpublished manuscript, taken from Ref.\onlinecite{pug89}, showing
		his construction of an ellipse whose vertices $\alpha$, $\beta$ and $\kappa$, etc.,
		were marked at half the distances between the central vertical and the circumscribing circle.
	    For a more detailed description, see Nauenberg\cite{nau94} and Erlichson\cite{erk97}.
}
	\label{hookorig}
\end{figure}

    This acrimony not only persisted in Hooke's and Newton's life time, 
but flared up dramatically in 1989 when Pugliese\cite{pug89} found, among Hooke's 
unpublished manuscripts, a drawing, Fig.\ref{hookorig},
dated September 1, 1685, 10 months after the {\it De Motu}, 
but nearly two years before the publication of
the {\it Principia},  seemingly showing an elliptical 
orbit due to a central linear (``Hooke's law") force! 
Nauenberg\cite{nau94,nau05} has seized upon
this diagram as the basis for reassessing Hooke's mathematical ``ability", 
arguing that Hooke has achieved something even Newton did not, that of using a
graphical method to solve an {\it inverse} problem of determining the orbit from a 
{\it given} force. A claim that has since repeated by Cooper\cite{coo03b}.

However, Nauenberg's conclusion was disputed by Erlichson\cite{erk97}, 
who argued that Hooke had simply constructed an ellipse 
{\it geometrically} by uniformly squeezing a circle 
and found that the resulting vertices can be fitted according to
Newton's graphical method for a linear force. Therefore, at most, what
Hooke has solved is the {\it direct} problem, that {\it given} a centered
elliptical orbit, the force can be determined to be linear.
 
   It is difficult even for experts to evaluate each side
of the argument. Clearly, Hooke has drawn a circumscribing circle in Fig.\ref{hookorig}. 
His orbital vertices $\alpha$, $\beta$, $\kappa$, etc., were marked precisely at midpoints\cite{nau94a} 
of chord lines drawn between the central vertical to points on the circle's circumference, 
a known method (an affine transformation) of constructing an ellipse. 
This fact was admitted as possible by Nauenberg in his original article\cite{nau94} 
and in his rebuttal\cite{nau98} to Erlichson. However, Hooke's text 
accompanying the diagram did mention that each impulse produced by the attractive 
body is proportional to the radial distance. 

A related puzzle is that if Hooke had  been successful in graphing the orbit of a 
linear force, why didn't he apply it to the all-important inverse-square force? 
Many of Hooke's supporters\cite{nau94,coo03,pur09} believed that he did,
but those drawings were simply lost. When referring to Fig.\ref{hookorig},
Purrington\cite{pur09a} ventured that ``It is limited to the case of a linear force, 
and while no similar calculation is known for the crucial case of the inverse-square force, 
such a proof would be not terribly more difficult, and it does not require a great leap 
of faith to suppose that such a proof by Hooke once existed." 

Fortunately, modern understanding of Newton's graphical construction as a {\it symplectic integrator} 
can now shed light on these issues. 
A symplectic integrator is a canonical transformation which seeks to integrate Hamilton's equation 
to obtain the system's coordinate and momentum as a function of time.  
The crucial point about a symplectic integrator is that, the trajectory it produces, 
while only approximate for the original Hamiltonian it intended to solve, is {\it exact} for 
a Hamiltonian which is the original Hamiltonian plus some error terms.
The key contribution of this work is to show that, when Newton's
graphical method is applied to a linear force, its corresponding Hamiltonian with error terms
remained harmonic and exactly solvable. This means that the trajectory produced by Newton's graphical
method can be determined {\it analytically}. By comparing this analytical
solution to Hooke's construction, one can decide whether Hooke has applied Newton's method and 
obtained the correct elliptical orbit. Similarly, one can repeat the analysis for the inverse-square 
force, and ascertain whether Newton's graphical method can produce an ellipse with the force at
one of its focus. In short, this work substantiated many of Erlichson's\cite{erk97} original criticisms
with precise analytical findings made possible by the modern development of symplectic integrators.

In the following, Sect.\ref{newton} reviews Newton's
graphical construction, explain how its geometric formulation is translated
into algebraic expressions, and why it is a symplectic integrator. 
Sect.\ref{ellor} gives the analytical form of the elliptical 
orbit produced by Newton's graphical method for a linear force. Sect.\ref{tech} compares
Newton's graphical construction of an upright symmetric orbit, similar to the one found in Hooke's 
manuscript. Sect.\ref{error} repeats the orbit construction for the inverse
square force. Sect.\ref{con} summaries key conclusions of this work.

\section {Newton's graphical construction as a symplectic integrator} 
\la{newton}

Newton's graphical construction, the celebrated Proposition 1, 
Theorem 1 of the {\it Principia} is shown in Fig.\ref{prop1}. 
In order to fully appreciate its relationship to symplectic integrators,
it is necessary to review details of its construction.
Newton stated that when a planet is initially moving at a given velocity, 
it will move from A to B in a given time interval. If there were no force acting on it, 
then by Newton's first law, it will continue to move in the same direction in the next
time interval from B to c, with distance Bc equal to AB. 
Triangles SAB and SBc (S=Sun) then have equal area; since their bases AB and Bc  
are of equal length, and both have the same apex S. 
However, if at B, the planet is acted on by a force impulse directed toward S along BS, 
instantaneously changing the planet's velocity so that it is moving along BC instead, then it will 
arrive at C. The position C (this is the key point), is along the
new direction BC such that cC is {\it parallel} to BS. Since cC is parallel 
to BS, triangles SBc and SBC have the same perpendicular height to their common
base SB. Therefore, their areas are equal. It then follows that the area SBC equals
area SAB, since both are equal to area SBc. Repeating this process for D, E, etc., shows that 
for any central force, the discrete trajectory ABCDE will sweep out 
equal areas SAB, SBC, SCD, SDE, etc., at equal
time intervals, thus proving Kepler's second law for any time interval. 

\begin{figure}
	\includegraphics[width=0.90\linewidth]{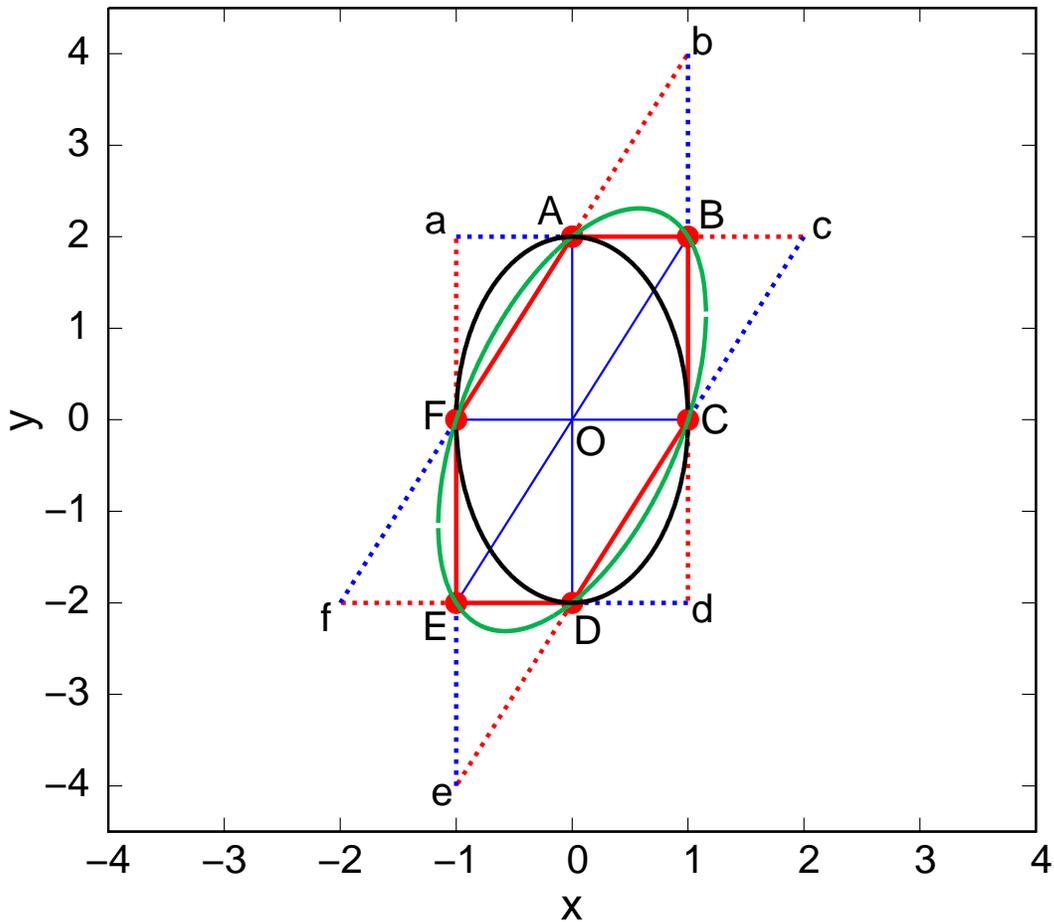}
	\caption{ Red dots are orbital positions produced by Newton's graphical method for a linear 
		central force at $\dt=1$. The green ellipse is its analytical description. The black ellipse 
		is the correct orbit. See text for details.  
	} 
	\la{orbit6}
\end{figure}

One can illustrate this construction concretely in Fig.\ref{orbit6} for a central linear force
with acceleration exactly equal to the radial distance.
One can take the initial position as ${\bf r_0}=(0,2)$, initial velocity ${\bf v_0}=(1,0)$
and time step $\dt=1$. Red dots A, B, C, D, E, and F are discrete positions
generated by the the analytical forms (\ref{first}) and (\ref{second}) 
given below, which anyone can plainly see as precisely matching
Newton's graphical construction as follow:
Starting at position A, the planet 
moves at constant velocity $v_{x0}=1$ for time $\dt=1$ to position B. If there were no 
force, the planet would have drifted at the same velocity to c in another $\dt$ time interval. 
However, at B, the central force exerted an impulse exactly equal to the distance BO, 
thereby changing velocity direction from AB to BC instantaneously. 
Because of this new velocity, the planet actually 
moved from B to C in time interval $\dt$. 
The position C is precisely along the new trajectory Bd such that cC is parallel to BO.
With the new velocity BC, if there were no force, 
the planet would again drift on from C to d. However, the impulse CO at C, 
changes the velocity BC to CD, and the planet actually arrived at D, and so on.
The unit triangular areas OAB, OBC, OCD, ODE, OEF, OFA,
sweep out by the polygonal trajectory ABCDEF, are obviously equal.  
The black and green ellipses will be discuss below.

Newton's graphical construction consisted of iterating two steps: 
1) Drifting at a constant velocity for time $\dt$ and 
2) Changing the velocity according to impulse $\ac\dt$ instantaneously. 
These are just Newton's first and second laws respectively. 
As first noted by Nauenberg\cite{nau94}, these two steps,
occurring at {\it one} time interval $\dt$, can be identified analytically as
\ba
\br'&=&\br+\bv\dt
\la{first}
\\
\bv'&=&\bv+\ac(\br')\dt.
\la{second}
\ea
For a central linear or inverse square force, 
$\ac(\br)=f(r)\br$, with $f(r)=-\omega^2$ or $f(r)=-k/r^3$ respectively.
Note that the updated position $\br'$ is used immediately in $\ac(\br')$
to update the velocity. This {\it sequential} updating is at the heart of 
Newton's graphical method, clearly shown in Fig.\ref{prop1},
but easily overlook. The force impulse is evaluated at B, {\it after} drifting from A,
and evaluated at C, {\it after} drifting from B, and so on.
Newton singles out the drift step (\ref{first}) first, then follow by the
change in velocity (\ref{second}) {\it together} with the next iteration of (\ref{first}).
That is, Newton iterates his algorithm at successive $\dt$ to the rhythm of
(1), (2)(1), (2)(1), etc., always 
ending with (1) drifting into the final position with the updated velocity. 
For example, if (2) were followed by by (1) again, one would get
\be
\br''=\br'+\bv'\dt=\br+\bv\dt+\bv\dt+(\ac(\br')\dt)\dt,
\ee
corresponding to the initial position $\br$ drifted $2\dt$ at the original velocity $\bv$, 
then the impulse (the change of velocity) $\ac(\br')\dt$ is evaluated at the previous position 
$\br'$, and drift for a time $\dt$ to the final position. 
This {\it precisely} corresponds to the initial drift from A to c in Fig.\ref{prop1},
then correcting the trajectory from c to C . Note that the trajectory correction (or deviation)
$cC=(\ac(\br')\dt)\dt$, is parallel to the impulse $\ac(\br')\dt$ evaluated 
at the {\it prior} position $\br'$. 

Equations (\ref{first}) and (\ref{second}) also gives an analytical proof of 
Kepler's equal area law. Referring to Fig.\ref{prop1}, let $SA=\br$ and $AB=\bv\dt$,
then area $SAB=|\br\times\bv\dt|/2$. After the updating according to 
(\ref{first}) and (\ref{second}), the area $SBC=|\br'\times\bv'\dt|/2$
will be equal to that of $SAB$, since
\be
\br'\times\bv'=\br'\times(\bv+f(r')\br'\dt)=(\br+\bv\dt)\times\bv=\br\times\bv,
\la{area}
\ee
independent of the actual form of $f(r)$. Thus, as it is well known, Kepler's equal area
law is just angular momentum conservation.
 
This equal area law is is not evident in Hooke's notion of ``compounding"
inertial and attractive motions. For example,
if one were to replace (2) by $\bv'=\bv+\ac(\br)\dt$, which evaluates the
force at the old position $\br$, and {\it not}
at the updated position $\br'$, then the result would be the {\it disastrous} 
Euler algorithm,\cite{chin16}
with no conservation of angular momentum, 
and for a non-constant force, unstable at any $\dt$, 
no matter how small. 
Yet, such a mathematical algorithm would still be consistent with Hooke's notion of 
``compounding" inertial motion with an attractive force.
Therefore, {\it for orbital motion around a central force}, what is needed,
but {\it missing} from Hooke's hypothesis, is a special way of ``compounding" that
would conserve angular momentum, {\it i.e.}, respecting Kepler's area law. 
Even {\it if} he had been influenced by Hooke's idea of ``compounding",
Newton's way of ``compounding", of incorporating Kepler's second law,
is entirely due to his own genius.  
 
Newton's graphical construction, as described earlier, and obvious from its
analytical form (\ref{first})-(\ref{second}), is that it is {\it sequential}.
As emphasized in Ref.\onlinecite{chin16}, this is a defining hallmark of being a
{\it symplectic integrator}. As a matter of fact, it is one of two fundamental,
long known first order symplectic integrator.\cite{chin16}
(The other one, Cromer's algorithm,\cite{cro81} is the sequential updating of
(2) followed by (1).) When Nauenberg\cite{nau94} first wrote down   
(\ref{first})-(\ref{second}), he did not know that they corresponded to 
a symplectic integrator.
Those who knew (\ref{first})-(\ref{second}) as a symplectic integrator 
were unaware of its connection to Newton's graphical construction. 
This work, aware of this connection, also knows the Hamiltonian which
govern the trajectories produced by (\ref{first})-(\ref{second}).
 
If the original motion is governed by the Hamiltonian 
$H=\p^2/2m+V(\br)$ then the trajectory generated by Newton's algorithm 
(\ref{first})-(\ref{second}) is governed 
by the approximate Hamiltonian,\cite{yos93}
\be
H_A=H-\frac12\p\cdot\ac(\br)\dt+O(\dt^2),
\la{herr}
\ee
where $\bv=\p/m$ and $\ac(\br)={\bf F}(\br)/m$. For the harmonic potential 
$V(\br)=m\w^2\br^2/2$, the above Hamiltonian with only the first order $\dt$ term,
is {\it exact} for Newton's algorithm. (All higher order $\dt$ terms only sum to an 
overall multiplicative constant\cite{scu05b,lar83}.) 
Thus, as will be shown in the next Section, for a linear central force,
the trajectory produced by Newton's graphical method is known {\it analytically}.

\section {Elliptical orbits of a linear force}
\la{ellor} 

Consider a central linear force in two dimension, the Hamiltonian with unit mass is
\be
H=\frac12 \left(\bv^2+\w^2\br^2\right),
\la{hor}
\ee  
with trajectory given by
\be
\br(t)=\br_0\cos(\w t)+\frac{\bv_0}{\w}\sin(\w t).
\la{traj}
\ee
For $\br_0=(0,y_0)$, $\bv_0=(v_{x0},0)$ one has
\be
x(t)=\frac{v_{x0}}{\w}\sin(\w t),\qquad y(t)=y_0\cos(\w t),
\ee
which is the elliptical orbit 
\be
\left(\frac{x}{v_{x0}/\w}\right)^2+\left(\frac{y}{y_0}\right)^2=1.
\la{orb}
\ee
For $\w=1$, $v_{x0}=1$ and $y_0=2$, this is the exact orbit
$x^2+(y/2)^2=1$, the black ellipse, 
shown in Fig.\ref{orbit6}.

The Hamiltonian (\ref{herr}) corresponding to Newton's algorithm  is
\ba
&&H_{A}=\frac12 (\bv^2+\w^2\br^2+\dt\w^2\bv\cdot\br)\nn\\
&&\ \ =\frac12 \left((\bv+\frac{\w^2\dt}2\br)^2+\w^2(1-\frac{\w^2\dt^2}{4})\br^2\right),
\la{h1b}
\ea
which can be rewritten as
\be
H_{A}=\frac12 \left(\widetilde\bv^2+\W^2\br^2\right),
\la{h1bp}
\ee
with $\widetilde\bv=\bv+\w^2\dt\br/2$ and $\W=\w\sqrt{1-\w^2\dt^2/4}$.
Therefore as long $|\dt|<2/\w$, this remains a harmonic oscillator with
angular frequency $\W$ and trajectory
\be
\br(t)=\br_0\cos(\W t)+\frac{\widetilde\bv_0}{\W}\sin(\W t),
\la{htraj}
\ee
where $\widetilde\bv_0=\bv_0+\w^2\dt\br_0/2$. Here, $H_{A}$ is the Hamiltonian exactly
conserved by Newton's algorithm (1)(2) in that $H_{A}(\br',\bv')=H_{A}(\br,\bv)$.

For a given $\dt$, and with the same initial conditions as
above, the trajectory generated by Newton's algorithm (\ref{htraj}) is
\ba
x(t)&=&\frac{v_{x0}}{\W}\sin(\W t),\nn\\ 
y(t)&=&y_0\cos(\W t)+\frac{\w^2\dt y_0}{2\W}\sin(\W t),
\ea
which is a closed elliptical orbit
\be
\left(\frac{x}{v_{x0}/\W}\right)^2+\left(\frac{y-(\frac{\w^2\dt y_0}{2v_{x0}})x}{y_0}\right)^2=1.
\la{orb2}
\ee
Thus Newton's graphical construction, when applied to a linear central force, will {\it always yield
orbital positions on a closed ellipse}, as long as $|\dt|<2/\w$. 

This surprisingly result is the key contribution this work. Normally, when one applies a numerical
method, one has no expectation that it will yield anything resembling the exact result at
a large value of $\dt$. To illustrate this we compare Newton's graphically constructed orbit in 
Fig.\ref{obs} with trajectories produce by other well-known algorithms. 

\begin{figure}
	\includegraphics[width=0.90\linewidth]{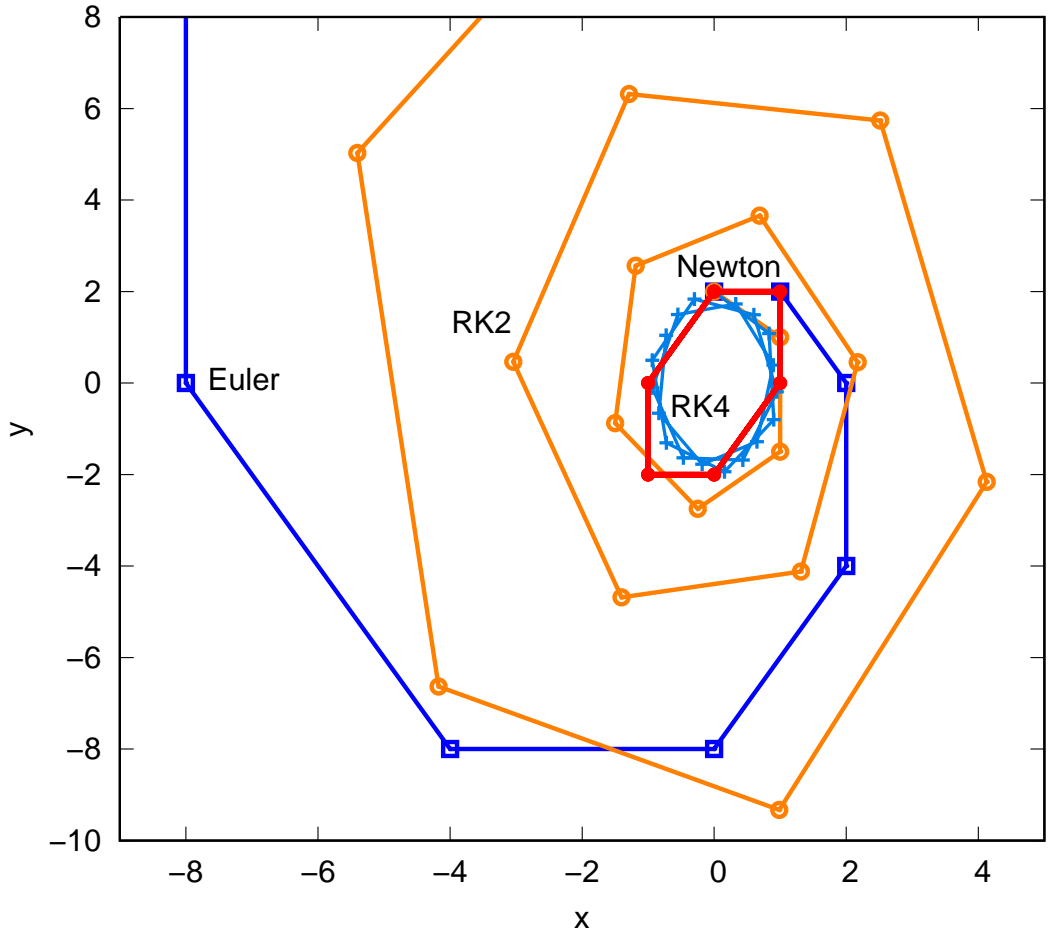}
	\caption{Comparing Newton's graphical method (red) with Euler's algorithm (blue), 
		second-order Runge-Kutta algorithm RK2 (orange) and fourth-order Runge-Kutta algorithm RK4 (light blue)
		at $\dt=1$ for 18 iterations. Only Newton's graphical construction remains a closed orbit.
	} 
	\la{obs}
\end{figure}

That's why Erlichson\cite{erk97} remarked that, had Hooke 
used Newton's method to solve the inverse problem, 
``then there would be no requirement that the vertex points lie on-orbit [on the elliptical orbit]".
He therefore thought that Hooke's results were too good to be true, and 
can't possibly from using Newton's method.
However, Nauenberg\cite{nau98}, also observed that ``In general, a graphical construction can give only an approximation to such a curve, but in this case, the resulting vertex points of the polygon lie on 
an ellipse." Since Hooke's construction is consistent with using a linear force, 
Nauenberg drawn the opposite conclusion that Hooke had produced an elliptical orbit using Newton's
graphical method. Both conclusions were wrong because both tacitly assumed that 
the ellipse drawn by Hooke is the correct orbit for the linear force. 
They did not anticipate the result of this work, that Newton's graphical method, 
when applied to a linear force, will always yield positions on a closed ellipse, 
but that ellipse is {\it not} the correct elliptical orbit! 

That ellipse, (\ref{orb2}), is shown in Fig.\ref{orbit6} as the green ellipse encompassing 
the polygonal orbit ABCDEF. 
This green ellipse approaches the correct black ellipse only in the limit of $\dt\rightarrow 0$. 
At any finite $\dt$, this ellipse has a characteristic ``tilt" from the vertical because its first step $\br_1=(v_{x0}\dt,y_0)$ is asymmetric with respect to the $y$-axis. 
Therefore, as originally noted by Erlichson\cite{erk97}, had Hooke just 
applied Newton's graphical method, using similar initial conditions, 
he would not have gotten an upright, left-right symmetric ellipse (to be discuss in the next Section). 
The reason why Hooke did not choose such a natural starting point is clear;
he did not apply Newton's graphical method.

\begin{figure}
	\includegraphics[width=0.90\linewidth]{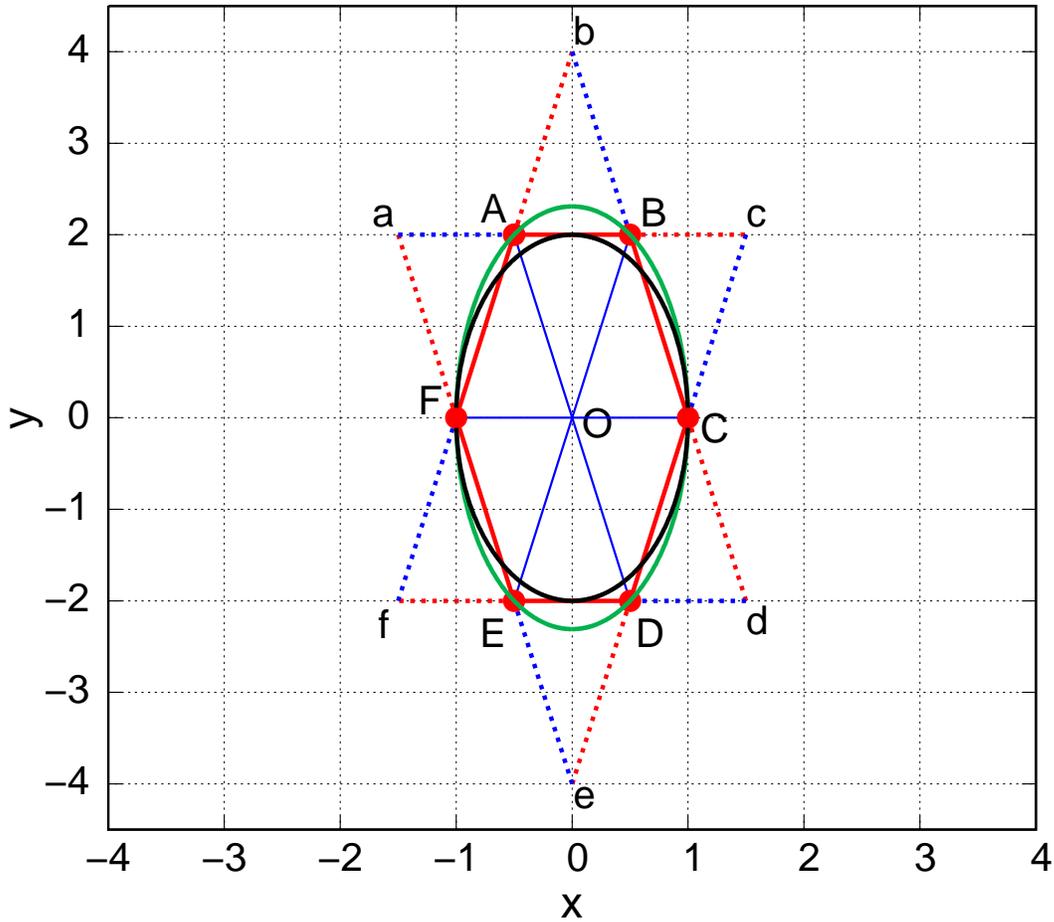}
	\caption{linear force orbit (red dots) generated by Newton's graphical method
		at $\dt=1$ with a pair of symmetric initial positions A and B. 
	} 
	\la{orbit62}
\end{figure} 

\section {The upright symmetric ellipse}
\la{tech}

Instead, Hooke chooses the initial position to be 
$\br_0=(-v_{x0}\dt/2,y_0)$, so that after step (1) of Newton's algorithm, $\br_1=(v_{x0}\dt/2,y_0)$. This then guarantees that the first two vertices of
the polygonal orbit will be
symmetric with respect to the $y$-axis. The resulting orbit from (\ref{htraj}) 
is then given by
\ba
\frac{\w^2 x(t)}{v_{x0}}&=&-\frac{\w^2\dt}2 \cos(\W t)+{\W}\sin(\W t),\nn\\
\W\frac{y(t)}{y_0}&=&\W\cos(\W t)+\frac{\w^2\dt}2 \sin(\W t).
\ea
The cross terms cancel when both sides are squared and added.
Since $\W^2+\w^4\dt^2/4=\w^2$, this gives remarkably, 
\be
\left(\frac{x}{v_{x0}/\w}\right)^2+\left(\frac{y}{y_0/\sqrt{1-\w^2\dt^2/4}}\right)^2=1,
\la{symorb}
\ee
which differs from the exact orbit of (\ref{orb}) only by the factor 
$(1-\w^2\dt^2/4)^{1/2}$!
Such a starting condition not only produces an upright symmetric ellipse, but 
also dramatically {\it improved} the algorithm from first to 
second order in $\dt$! There is now no ``tilt" proportional to $\dt$.
Using the same parameter values as in Fig.\ref{orbit6}, the algorithm's
orbit (\ref{symorb}) is shown as the green ellipse ABCDEF in 
Fig.\ref{orbit62}. This green ellipse is now much closer 
to the black ellipse, but it is still not the correct orbit!  
The discrete graphical constructed points 
are again A, B, C, D, E and F. It is now even more obvious that Kepler's equal 
area law is obeyed: the areas swept out are equal area segments of the hexagon ABCDEF.

\begin{figure}
	\includegraphics[width=0.90\linewidth]{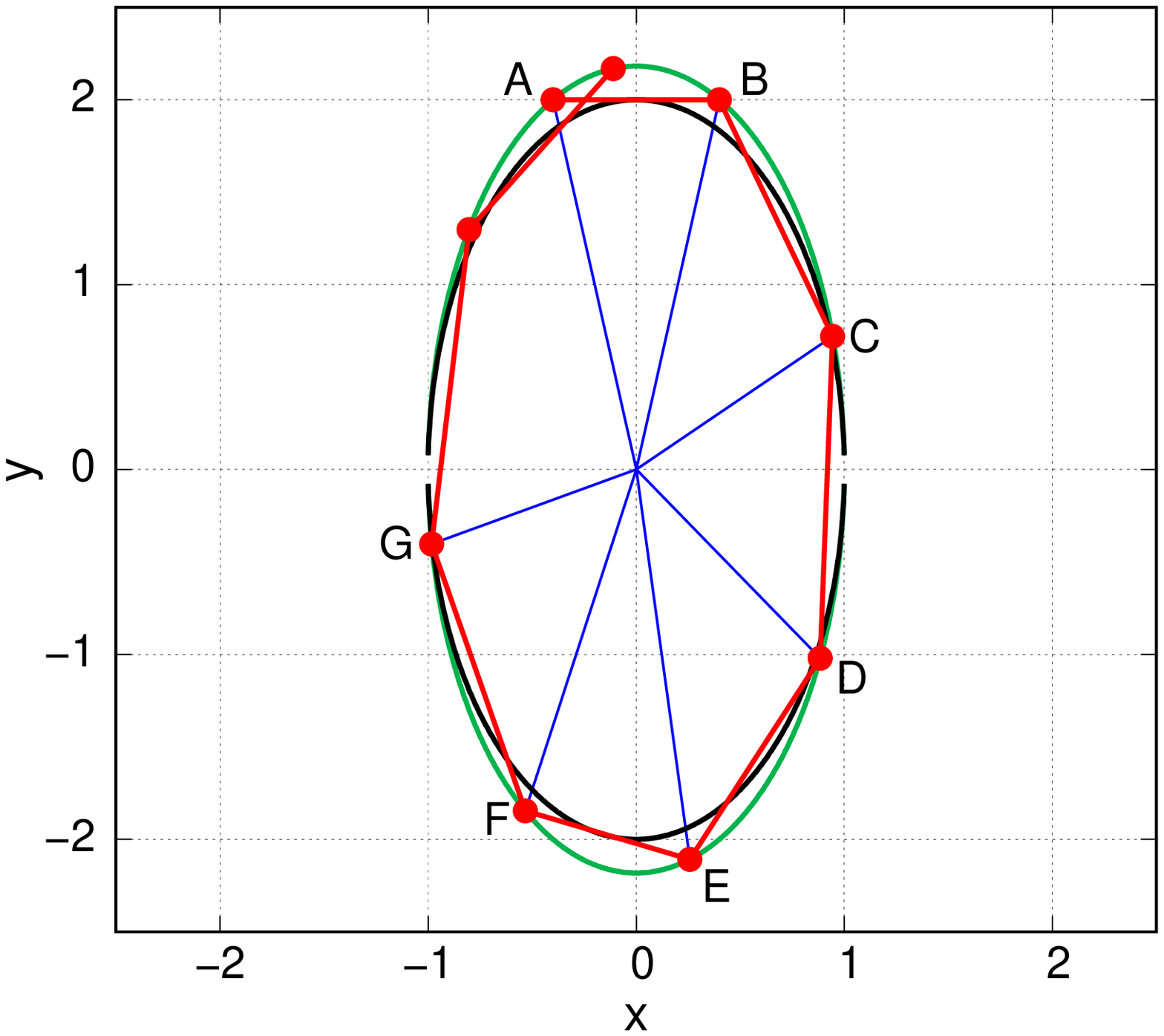}
	\caption{Same as Fig.\ref{orbit62} but with $\dt=0.8$. 
	} 
	\la{orbinc}
\end{figure}

However, the choice of $\dt=1$, which divided the algorithm's orbit exactly into six segments,
is rather special. Although the orbit of the algorithm is determined by $\W$, the actual angular
velocity of the algorithm is given by $\W_A$, defined by\cite{scu05b}
\be
\cos(\W_A\dt)=1-\w^2\dt^2/2.
\la{wa}
\ee 
Therefore, if one were to chose a $\dt$ such that $T/\dt=n$, where the period $T=2\pi/\W_A$,
then one must have $\W_A\dt=2\pi/n$ or
\be
\dt=\frac1{\w} \sqrt{2\Bigl(1-\cos\Bigl[\frac{2\pi}n\Bigr]\Bigr)},
\la{wn}
\ee
which gives $\dt=1$ for $\w=1$ and $n=6$. For any other arbitrary value of $\dt$, say $\dt=0.8$, $\dt$ would not divide the period evenly. This is shown in Fig.\ref{orbinc}. The positions produced by the graphical method still lie on the predicted green ellipse (\ref{symorb}), but are no longer left-right symmetric. This is another of Erlichson's criticism.\cite{erk97} Therefore, if Hooke were simply applying Newton's graphical method, he would have gotten, either the tilted polygonal orbit of Fig.\ref{orbit6}, or an upright but left-right {\it asymmetric}
polygonal orbit of Fig.\ref{orbinc}. The fact that Hooke is marking off parallel chord lines in Fig.\ref{hookorig} shows that he is using the circumscribing circle to {\it construct} 
left-right symmetric elliptical vertices\cite{erk97}. 
But such vertices are highly unlikely, because unless $\dt$ is specifically chosen 
according to (\ref{wn}), the resulting orbital vertices {\it cannot} be left-right symmetric. 

Nauenberg\cite{nau94} also claimed that Hooke's diagram (Fig.\ref{hookorig}) 
proved the area law differently from Newton.
Newton's graphical proof of Kepler's area law is general, true for {\it any} central force.
Hooke's demonstration of the area law is {\it specific} to his ellipse construction. 
This is because Hooke is uniformly squeezing the circle; 
areas of pie-like slices on the circumscribing circle are all
squeezed by the same factor into areas of pie-like slices on the ellipse. 
Therefore, if the original slices on the circle are of equal area, then the
squeezed slices on the ellipse are also of equal area. This is just a geometric
consequence of Hooke's construction, not a general embodiment of Kepler's equal area law.

\section {The inverse-square force}
\la{error}

What happens when Newton's graphical method is applied to the gravitational force?
Why didn't Newton use it to compute Kepler's orbit? Since Newton has claimed 
in Corollary 1 of Book 1's Proposition 13, that an inverse-square force would yield 
an elliptical orbit, he would not have felt the need to use his graphical method to
prove anything else except Kepler's second law. However, others, such as 
Weinstock\cite{wei82}, have argued that Newton's Corollary 1 is not a proof,
and that Newton never solved the inverse problem for the inverse square force 
in the {\it Principia}. But is it possible that Hooke could have used the
same graphical method to produce an orbit for the inverse-square force, as suggested by
Purrington\cite{pur09a}?

\begin{figure}
	\includegraphics[width=0.90\linewidth]{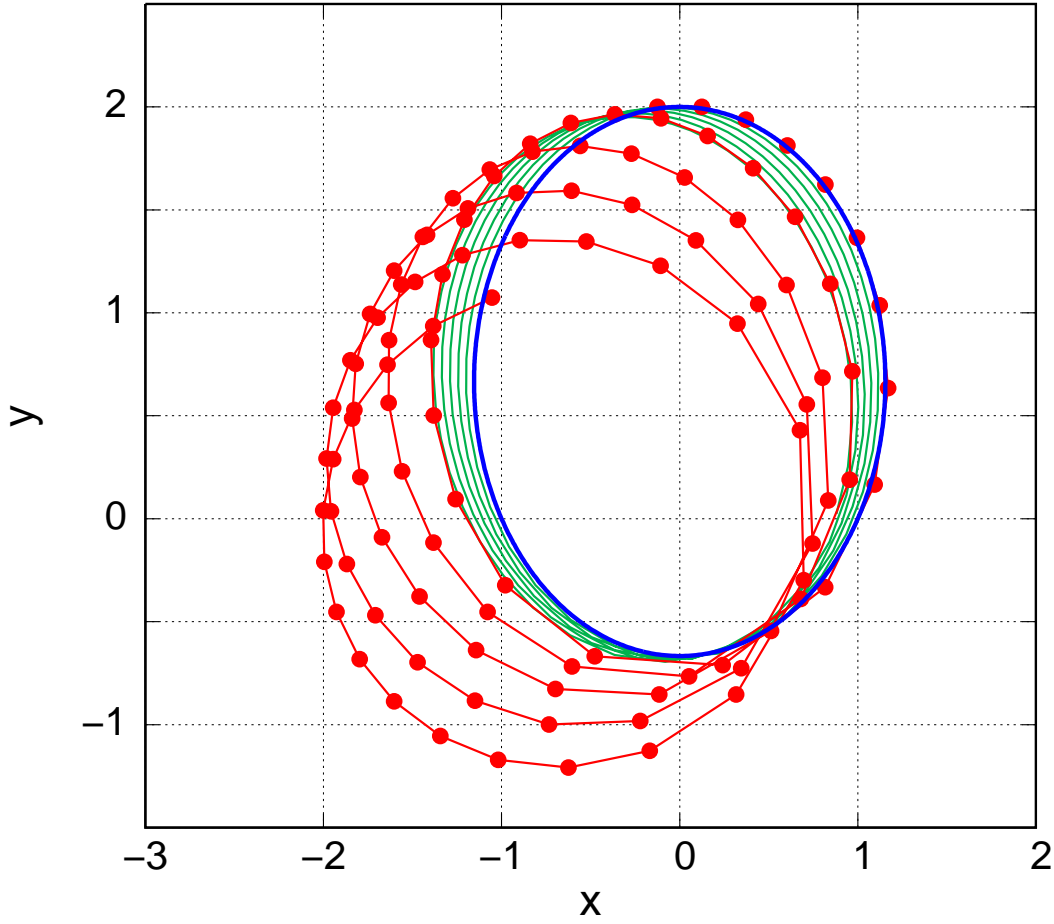}
	\caption{Inverse-square force orbits generated by Newton's graphical method at $\dt=0.5$ (red) $\dt=0.2$ (blue-green) and $\dt=0.01$ (blue). See text for details.
	} 
	\la{overr}
\end{figure} 

Nauenberg\cite{nau94} has shown that if $\dt$ is too large, the orbit will not be elliptical 
near the force center. Here, we can again study the trajectory produced Newton's graphical method
in greater details by analyzing the approximate Hamiltonian (\ref{herr}). 
For the inverse-square force, the approximate Hamiltonian is
\be
H_A=\frac12\bv^2-\frac{k}{r}+\frac{\dt}2 \frac{k\br\cdot\bv}{r^3}+O(\dt^2).
\la{hinv}
\ee
In the linear force case, the 
first-order error term ($\propto \bv\cdot\br$) is {\it less} singular 
(as $\br\rightarrow\infty$) than
the original potential ($\propto \br^2$), and therefore the approximate Hamiltonian 
can still be redefined as harmonic (\ref{h1bp}) with a potential $\propto \br^2$. 
This then produces at most, a tilted, but still {\it closed} elliptical orbit (\ref{orb2}).
By contrast, for the gravitational force, the first-order $\dt$ term 
above ($\propto \bv\cdot\br/r^3$) is {\it more} singular (as $\br\rightarrow 0$) than the original 
potential ($\propto 1/r$) and can never be redefined so that the approximate Hamiltonian is having
a potential $\propto 1/r$. By Bertrand's theorem\cite{ber73},
if the potential is neither $r^2$ nor $1/r$ {\it exactly},
then the orbit cannot be closed. Thus when applied to the gravitational force, 
Newton's graphical method, even if $\dt$ is sufficiently small so that 
the trajectory can go around the force center, {\it will not yield a closed orbit}.
The orbit will remain open and and just precess. 
For the kind of geometrical construction 
that can be done by hand, one would not get a recognizable ellipse. This is illustrated in Fig.\ref{overr} for $\dt=0.5$, $k=1$ and a symmetrical start with initial velocity reduced to $v_{x0}=0.5$. The exact orbit with eccentricity 1/2 given by
\be
\left(\frac{x}{2/\sqrt{3}}\right)^2+\left(\frac{y-2/3}{4/3}\right)^2=1,
\ee 
can be produced by Newton's algorithm (1)-(2) with $\dt=0.01$. The red dot trajectory generated
by $\dt=0.5$ also coincide with the correct orbit as it proceeds down the right side of
the ellipse. However, as noted by Nauenberg\cite{nau94}, this is no longer the case
after the trajectory passes below the force center at the origin. 
Further iterations show that the orbit precesses counter-clockwise. 
After $\approx$5 periods, the semi-major axis has 
rotated 90$^\circ$. (The precession shown is actually due to the 
$\dt^2$ terms (not shown) in (\ref{hinv}) because the $\dt$ term contribution vanishes after
each period. To verify this $\dt^2$-dependence, Fig.\ref{overr} also shows that the orbital 
precession for $\dt=0.2$ is roughly $(0.5/.2)^2\approx 6$ times smaller than that of $\dt=0.5$.)

Thus when Newton's graphical method is applied, the resulting trajectory differs greatly 
depending on whether the central force is linear or inverse-square. For the inverse square force
a closed orbit is possible only for very small time steps, such as $\dt=0.01$, 
as shown in Fig.\ref{overr}. However, such a time step is 
probably 50 times too small to be drawn by hand. Most likely, a hand construction would
resemble the red dot orbit in Fig.\ref{overr}, corresponding to $\dt=0.5$.
In this case, the orbit will not close. Therefore, contrary to Purrington's musing,
it is also unlikely that Hooke could have constructed by hand,
a closed elliptical orbit for the inverse square force.  

\section {Conclusions}
\la{con}

This work, by knowing the Hamiltonian which governs the trajectory of a symplectic
integrator, has shown that when Newton's graphical method is applied to a central
linear force, the resulting trajectory is always an centered ellipse, 
just not the correct elliptical orbit.
By carefully comparing this analytically ellipse to Hooke's drawing,
this work substantiated Erlichson's conclusion that Hooke was trying to repeat what he 
had done before with a circular orbit, that he was trying to follow an elliptical orbit in order to
determine the required force. He constructed an elliptical orbit, then found that,
consistent with Newton's graphical construction, the required force is linear.  
This solves the direct, not the inverse problem. Moreover, even this success is
fortuitous; because by the same result above, it is a peculiar property of 
Newton's graphical method that any centered ellipse can be fitted to a linear force.

When the same analysis is applied to the inverse square force,
the result is drastically different. When Newton's graphical method is applied by hand to
the inverse square force, it is unlikely to yield a closed, elliptical orbit.
This is the more likely reason why no similar drawing for the inverse square force
can be found in Hooke's possession.

\begin{acknowledgments}
I thank my colleague Wayne Saslow, for initiated my interest in this work.
\end{acknowledgments}


\end{document}